\documentclass{article}
\usepackage{arxiv}

\usepackage{amsmath,amsfonts,amssymb}
\usepackage{mathtools}
\usepackage[hidelinks]{hyperref}
\usepackage{algorithmic}
\usepackage{algorithm}
\usepackage{array}
\usepackage[caption=false,font=normalsize,labelfont=sf,textfont=sf]{subfig}
\usepackage{relsize}
\usepackage{textcomp}
\usepackage{stfloats}
\usepackage{url}
\usepackage{verbatim}
\usepackage{graphicx}
\usepackage{natbib}
\usepackage{doi}
\usepackage{xcolor}
\usepackage{paralist}
\usepackage{blindtext}
\usepackage{multicol}
\usepackage{multirow}
\usepackage[subnum]{cases}

\newtheorem{theorem}{Theorem}[section]

\newtheorem{remark}[theorem]{Remark}

\title{Socially Compliant Control of Autonomous Vehicles with Application to Eco-Driving}

\author{Shian Wang\\
	Electrical and Computer Engineering\\
	The University of Texas at El Paso\\
	\texttt{swang14@utep.edu} \\
}

\renewcommand{\shorttitle}{Socially Compliant Eco-Driving Control of Autonomous Vehicles}

\begin{document}
\maketitle

\begin{abstract}
Control design of autonomous vehicles (AVs) has mostly focused on achieving a prespecified goal for an individually controlled AV or for a swarm of cooperatively controlled AVs. However, the impact of autonomous driving on human-driven vehicles (HVs) has been largely ignored in AV controller synthesis, which could result in egoistic AV behavior detrimental to the safety of passengers and surrounding traffic. In this study we develop a general framework for socially compliant control design of AVs with a useful metric of social psychology, called social value orientation (SVO), allowing AVs to leverage their impact on the behavior of the following HVs. This is critical since AVs that behave in a socially compliant manner enable human drivers to comprehend their actions and respond appropriately. Within the proposed framework, we define the utilities of the controlled AV and its following vehicle, to be maximized in a weighted fashion determined by the AV's SVO. The utility maximization covers an array of design objectives given the goal of the AV and the benefits for the following HV stemming from the courtesy of socially compliant AV controls. An optimal control problem is then formulated to maximize the utility function defined, which is numerically solved using Pontryagin's minimum principle with optimality guarantees. The methodology developed is applied to synthesize socially compliant control for eco-driving of AVs. A set of numerical results are presented to show the mechanism and effectiveness of the proposed approach using real-world experimental data collected on Highway~55 in Minnesota.
\end{abstract}


\section{Introduction}\label{section1}


With the advancement of sensing, perception, and automation technologies, autonomous agents are becoming increasingly capable of accomplishing complex tasks in highly dynamic environments. For example, specialized robots have been proven successful in industrial tasks, hazardous environments, and medical application, among others~\citep{sheridan2016human}. The real-world environment in which autonomous agents are operated is intrinsically complex with various levels of uncertainty. As a result, efficient design of control and operation strategies for those robotic agents requires a number of interdisciplinary contributions involving human factors, engineering, cognitive psychology, and so on~\citep{goodrich2008human}. For instance, a cognitive model has been applied to help simulate the human problem-solving process leading to improved human cognition techniques, which further enables effective human-robot interactions~\citep{farouk2022studying}. Essentially, desired human-robot interactions allow autonomous agents to adapt to individual preferences and adjust their behavior according to the dynamic environment they are in~\citep{mitsunaga2008adapting}. By contrast, operating autonomous agents without considering their interaction with and impact on the surrounding environment including human operators could degrade the overall performance of the system, and may even result in destructive consequences. 

An important application of human-robot systems is autonomous driving in mixed traffic flow consisting of autonomous vehicles (AVs) and human-driven vehicles (HVs). AVs are expected to drastically revolutionize future transportation systems, offering a multitude of advantages, including reduced energy consumption~\citep{sun2022energy}, enhanced traffic stability~\citep{wang2022optimal}, and efficient utilization of parking spaces~\citep{wang2021optimal}, among others. However, to achieve many of the aforementioned benefits requires a fairly high penetration rate of AVs in the flow~\citep{fagnant2015preparing,seo2017endogenous}. The ground transportation system is expected to witness a coexistence of AVs and HVs for the foreseeable future as it transitions towards full automation~\citep{wang2022policy}. Consequently, human-AV interactions are expected to play an important role in affecting traffic performance while the system transitions from being partially automated to becoming fully automated. 

To achieve the full potential of AVs one needs to synthesize appropriate AV controllers depending on the performance goals. For example, a series of AV control strategies have been developed for efficient eco-driving controls~\citep{sun2022energy,wang2019cooperative}, autonomous intersection management~\citep{levin2017conflict,wang2023optimal}, and traffic smoothing~\citep{wang2022optimal,wang2023general}, considering either fully or partially automated traffic flow. While the AV control policies proposed in those studies could improve the performance of AVs in certain aspects, they have hardly taken into account the interactions between AVs and the surrounding HVs in the control design process. However, considering human-AV interactions in designing AV controllers is critical for ensuring the safety of passengers and the surrounding traffic; AVs that behave in a socially compliant manner enable human drivers to comprehend their actions and respond appropriately~\citep{schwarting2019social}. Such interactions could be even more pivotal to consider when it comes to vulnerable road users as safety is the top priority of transportation systems~\citep{reyes2022vulnerable}. 

Recently, a useful concept of social interactions for autonomous driving has been proposed to inform the design of socially compliant AV controllers~\citep{wang2022social}. For instance, a cognitive model, i.e., a type of trust models, has been constructed for a human-vehicle collaboration considering mutual trust~\citep{basu2016trust}. Due to the inevitable interactions between an AV and its surrounding traffic environment, it is believed that AVs may degrade the performance of the transportation system without considering human-AV interactions in designing AV controls~\citep{jafary2018survey}. Some notable and relevant studies on socially compliant control design of AVs have been presented in~\citep{buckman2019sharing,buckman2021semi,ozkan2021socially}. Specifically, a coordination policy that swaps AV ordering to improve the overall efficiency has been developed for autonomous intersection management while ensuring that the swaps are socially compliant~\citep{buckman2019sharing}, without having to strictly follow the first-come, first-serve ordering~\citep{levin2017conflict}. This is further extended to semi-cooperative control of an autonomous emergency vehicle using a game-theoretic approach for efficient interactions with HVs on the road~\citep{buckman2021semi}. In addition, a socially compatible control design is developed for AVs in mixed traffic with human-AV interactions incorporated into their decision-making process~\citep{ozkan2021socially}, which however requires preview information ahead of the AV assumed to be known via connectivity and advanced sensing. 

In contrast to prior work, in this study we focus on developing a general framework for AV control design with integration of a social psychology metric, called social value orientation (SVO)~\citep{liebrand1988ring,mcclintock1989social}, allowing AV controllers to leverage their impact on the behavior of the following HVs. Within the proposed framework, we design socially compliant control of AVs by maximizing the utilities of the AV and its following HV weighted by the AV's SVO. Notably, this framework covers an array of design objectives considering the practical desires of the AV as well as the benefits for the following HV due to the courtesy of socially compliant control of the AV. The utility maximization is properly formulated as an optimal control problem which is solved using Pontryagin's minimum principle~\citep{pontryagin1962mathematical} with optimality guarantees. As an example, we apply the methodology developed to synthesize socially compliant control for eco-driving of AVs. A set of numerical simulations are carried out to show the mechanism and effectiveness of the proposed approach using real-world vehicle trajectory data acquired on Highway~55 in Minnesota. The main contributions of this work are summarized as follows:
\begin{itemize}
    \item A general framework is developed for socially compliant AV control design based on social value orientation (SVO), allowing for inclusion of a range of design objectives depending on practical considerations. 
    
    \item An optimal control problem is formulated to maximize the utility of the AV and that of its following HV thanks to the courtesy of socially compliant control of the AV. The utility maximization problem is solved using Pontryagin's minimum principle with optimality guarantees. 
        
    \item The proposed methodology is applied to derive socially compliant control for eco-driving of AVs. A set of simulation results are presented to show the mechanism and effectiveness of the proposed approach using real-world experimental data collected on Highway~55 in Minnesota.
\end{itemize}

The remainder of this article is outlined as follows. In Section~\ref{section2}, we present a mathematical model for describing mixed traffic flow dynamics involving AVs and HVs. In Section~\ref{section3}, we first introduce social value orientation (SVO) and its integration into socially compliant control design of AVs. We then formulate an optimal control problem based on the mixed traffic dynamics and SVO presented in Section~\ref{section2} and Section~\ref{section3}, respectively, for deriving socially compliant eco-driving control of AVs. Further, a numerical solution algorithm is developed based on Pontryagin's minimum principle for solving the optimization problem formulated. In Section~\ref{section4}, we illustrate the mechanism and effectiveness of the proposed methodology via numerical simulation using real-world experimental data. The article concludes in Section~\ref{section5}, where future extensions are discussed.

\section{Mathematical Model of Mixed Traffic}\label{section2}

\begin{figure}[t!]
	\centering
	\includegraphics[width=\textwidth]{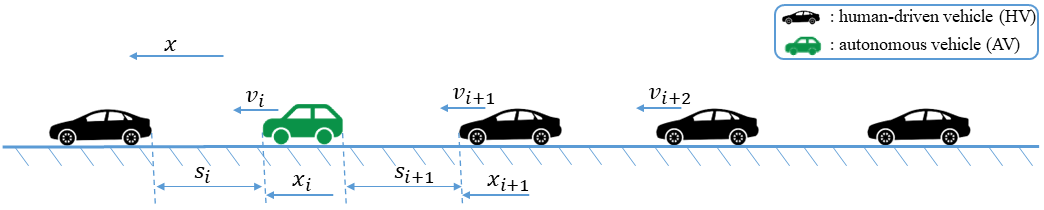}
	\caption{Graphic illustration of human-AV interactions in mixed traffic consisting of HVs and AVs, where $i \in \mathcal{A}$ and $(i+1) \in \mathcal{H}$}
	\label{mixed_traffic}
\end{figure}

In this study we focus on mixed traffic consisting of HVs and AVs shown in Fig.~\ref{mixed_traffic}, which is projected to be the case in the foreseeable future~\citep{wang2022policy}. Similar to many prior studies~\citep{wang2022optimal,wang2023general,talebpour2016influence}, we limit our focus to the longitudinal vehicle dynamics, recognizing that lateral dynamics could also be explored. We consider a general string of $n$ vehicles represented by the ordered set $\mathcal{N} = \left\{ 1, 2, 3, \cdots, n \right\}$, with $1 < n \in \mathbb{N}^{+}$. We signify the position and speed of vehicle $i \in \mathcal{N}$ at time $t$ as $x_{i}(t)$ and $v_{i}(t)$, respectively. The inter-vehicle spacing between vehicles $i$ and $i-1$ is defined as $s_{i}(t) = x_{i-1}(t) - x_{i}(t) - l_{i-1}$, where $l_{i-1}$ denotes the ($i-1$)th vehicle length. These notations have been widely employed in the description of car-following dynamics~\citep{wilson2011car}.

Based on car-following dynamics characterizing how vehicles follow one another on a roadway, the motion of any vehicle $i \in \cal{N}$ is given by~\citep{wilson2011car}
\begin{eqnarray}
    &~&\dot{x}_i(t) = v_i(t),    \label{eq2.1}   \\
    &~&\dot{v}_i(t) = f(s_i(t), \Delta v_i(t), v_i(t)),     \label{eq2.2}
\end{eqnarray}
where the dot operator signifies differentiation with respect to time; the operator $f$ connects the acceleration of the $i$th vehicle, $\dot{v}_{i}$, with the variables $s_{i}$, $\Delta v_{i}$ and $v_{i}$; and the definition of the relative speed between vehicle $i$ and vehicle $i-1$ is
\begin{eqnarray}
    \Delta v_i(t) = \dot{s}_i(t) = v_{i-1}(t) - v_i(t).     \label{eq2.3}
\end{eqnarray}
Equations~\eqref{eq2.1}--\eqref{eq2.2} embody a widely adopted functional expression for car-following dynamics. Although the variables, like $s_{i}$, $\Delta v_{i}$ and $v_{i}$, are time-dependent, we will, for the sake of brevity, omit the argument $t$ when applicable.

Let $\mathcal{H}$ and $\mathcal{A}$ represent the ordered sets of HVs and AVs within the composite traffic, respectively. As human drivers often demonstrate distinct driving behaviors compared to AVs~\citep{schwarting2019social}, we further differentiate $f$ in equation~\eqref{eq2.2} as $f_{\text{HV}}$ and $f_{\text{AV}}$ for HVs and AVs, respectively. For AVs, our goal is to synthesize appropriate additive controllers in a socially compliant manner applicable for a wide range of control objectives. To this end, the mixed traffic dynamics is written as
\begin{eqnarray}
    \dot{x}_i = v_i, ~ i \in \mathcal{N} = \mathcal{H} \cup \mathcal{A}   \label{eq2.4}
\end{eqnarray}
\begin{numcases}{\dot{v}_i =}
    f_{\text{HV}}(s_i, \Delta v_i, v_i), ~ i \in \mathcal{H}   \label{eq2.5a}  \\  [3pt]
    f_{\text{AV}}(s_i, \Delta v_i, v_i) + u_i, ~ i \in \mathcal{A}    \label{eq2.5b}
\end{numcases}
where $u_i$ represents the additive acceleration control input for an AV, designed to maintain adherence to car-following principles, as seen in~\citep{wang2022optimal,wang2023general,wang2022smoothing,delle2022new}. The acceleration that an AV executes is defined by the entire equation~\eqref{eq2.5b}. Within this equation, a specific component, namely the additive term $u_i$, requires explicit characterization. It is noted that the functionals $f_{\text{HV}}$ and $f_{\text{AV}}$ do not have to be the same since AVs often demonstrate distinct driving behaviors compared to human drivers~\citep{schwarting2019social}; instead they are contingent upon the particular car-following principles adhered to by HVs and AVs, respectively. 

In this study, we adopt the widely used intelligent driver model (IDM)~\citep{treiber2000congested} to describe the explicit dynamics of HVs, as seen in many previous studies~\citep{wang2023general,talebpour2016influence,shang2023extending,treiber2013traffic}. The IDM is a multi-regime model capable of offering a high level of realism in representing various congestion levels~\citep{sarker2019review}. Furthermore, recent research has demonstrated the IDM's ability to closely emulate human driving behavior with exceptional accuracy, outperforming other car-following models in real-world driving data assessments~\citep{pourabdollah2017calibration,he2023calibrating}.

Following the IDM, equation~\eqref{eq2.2}, or equation~\eqref{eq2.5a}, is explicitly written as
\begin{eqnarray}
    \dot{v}_i = a \left[ 1 - \left(\frac{v_i}{v_0}\right)^4  - \left(\frac{s^{\ast}(v_i,\Delta v_i)}{s_i}\right)^2 \right], ~i \in \mathcal{H},   \label{eq2.6}
\end{eqnarray}
with
\begin{eqnarray}
    s^{\ast}(v_i,\Delta v_i) = s_0 + \tau_1 v_i - \frac{v_i\Delta v_i}{2\sqrt{ab}},   \label{eq2.7}
\end{eqnarray}
where $a$ is the maximum acceleration, $b$ represents the comfortable braking deceleration, $v_{0}$ signifies the desired speed, $s_{0}$ denotes the minimum spacing, and $\tau_1$ indicates the desired time headway.

As noted in~\citep{schwarting2019social}, AVs often exhibit distinct driving behaviors compared to human drivers. While AVs from different manufacturers may feature varying acceleration functions described by $f_{\text{AV}}$, we adopt the optimal velocity with relative velocity (OVRV) model~\citep{milanes2013cooperative} to describe their car-following behavior. The OVRV model has been extensively used for (cooperative) adaptive cruise control (ACC) systems~\citep{liang1999optimal,milanes2013cooperative,gunter2019model}, i.e., the first generation of AVs. It adheres to a constant time-gap policy, which aligns with the implementation of intelligent vehicles~\citep{ioannou1993autonomous}. Furthermore, the OVRV model has demonstrated its capability to accurately represent both simulated and actual vehicle trajectories involving ACC vehicles~\citep{liang1999optimal,milanes2013cooperative}.

According to the OVRV model, equation~\eqref{eq2.2}, or $f_{\text{AV}}$ of equation~\eqref{eq2.5b}, is written as
\begin{eqnarray}
\dot{v}_i = k_1(x_{i-1} - x_i - l_{i-1} - \eta - \tau_2 v_i) + k_2(v_{i-1} - v_i), ~ i \in \mathcal{A},       \label{eq2.8}
\end{eqnarray}
where $\eta$ denotes the jam distance, i.e., inter-vehicle spacing at rest, $\tau_2$ represents the desired time gap, and $k_{1}$ and $k_{2}$ are positive parameters on the time gap and relative speed, respectively. Consequently, following equation~\eqref{eq2.5b} the AV dynamics with an additive control input is given by
\begin{eqnarray}
\dot{v}_i = k_1(x_{i-1} - x_i - l_{i-1} - \eta - \tau_2 v_i) + k_2(v_{i-1} - v_i) + u_i, ~ i \in \mathcal{A},     \label{eq2.9}
\end{eqnarray}
where the input $u_i$ is subject to the physical constraints $u_{\text{min}} \leq u_i \leq u_{\text{max}}$, with $u_{\text{min}}$ and $u_{\text{max}}$ being its lower and upper bounds, respectively.

\section{Socially Compliant Human-Robot Interactions with Application to Eco-driving of AVs}\label{section3}

In this section, we first introduce the concept of social value orientation (SVO) which is adopted to characterize socially compliant human-robot interactions, i.e., the interactions between HVs and AVs. We then formulate a general optimal control problem covering a broad range of objectives for AV control design. A concrete example is given for socially compliant eco-driving controls of AVs.

\subsection{Social Value Orientation (SVO)}\label{section3.1}

Social value orientation (SVO) is a useful metric from psychology that can be used to quantify social preferences of human behavior and their corresponding levels of cooperation~\citep{schwarting2019social,liebrand1988ring,mcclintock1989social,pletzer2018social}. SVO quantifies how an individual weights their payoff in relation to the payoffs of others, mainly translating into egoistic, prosocial, or altruistic preferences. Designing AVs that maximize their own payoff only may fail to account for nuances in real human behavior~\citep{schwarting2019social}, thereby adversely impact future traffic flow. Since SVO indicates a person's or an agent's inclination for distributing payoffs between themselves and another individual or agent, it can be used to effectively describe cooperative motives, negotiation strategies, and choice behavior~\citep{schwarting2019social,mcclintock1989social,pletzer2018social}. One can represent SVO preferences with a discrete-form triple dominance measure~\citep{van1997development}, a slider measure~\citep{murphy2011measuring}, or an angle $\varphi$ within a ring~\citep{liebrand1988ring}. Following the prior work~\citep{schwarting2019social}, in this study we adopt angular notation for SVO, shown in Fig.~\ref{SVO}.

One can set the AV's SVO for the design of socially compliant autonomous driving. This is critical for the safety of passengers and surrounding traffic, since AVs that behave in a socially compliant manner enable human drivers to comprehend their actions and respond appropriately~\citep{schwarting2019social}. To this end, we define as follows a utility function $J_1$ combining the payoff of the ego AV with that of its following vehicle, weighted by the AV's SVO angular preference $\varphi$
\begin{eqnarray}
    J_1 = \cos{(\varphi_{\text{AV}})} U_{\text{AV}} + \sin{(\varphi_{\text{AV}})} U_{\text{follower}},    \label{eq3.1}
\end{eqnarray}
where $U_{\text{AV}}$ and $U_{\text{follower}}$ represent the ``payoff to self'' and ``payoff to the following vehicle,'' respectively, and $\varphi_{\text{AV}}$ is the SVO of the AV. Clearly, the orientation of $\varphi_{\text{AV}}$ weights the payoff $U_{\text{AV}}$ against $U_{\text{follower}}$ according to the AV's actions. The subsequent definitions of social preferences are given based on these weights~\citep{schwarting2019social,liebrand1988ring,murphy2011measuring}:

\begin{figure}[t!]
	\centering
	\includegraphics[width=0.3\textwidth]{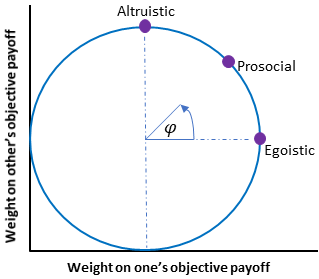}
	\caption{Graphic illustration of social value orientation (SVO) in angular notation, where $\varphi \approx \pi/2$, $\pi/4$, and $0$ correspond to `altruistic', `prosocial', and `egoistic', respectively.}
	\label{SVO}
\end{figure}

(1) Altruistic: Altruistic agents aim to maximize the outcome for the other party, without regard for their own payoff, with $\varphi \approx \pi/2$.

(2) Prosocial: Prosocial agents intend to benefit the collective welfare of a group, with $\varphi \approx \pi/4$. This is typically achieved by maximizing the joint payoff.

(3) Egoistic/individualistic: Egoistic agents prioritize maximizing their own payoff and do not take into account the outcome of the other party, with $\varphi \approx 0$. 


Following~\citep{schwarting2019social}, the definitions are limited to rational social preferences, and more details can be found in~\citep{liebrand1988ring,murphy2011measuring}. For clarity these definitions provide specific values for SVO preferences, however, it is noted that SVO exists in a continuous space. For instance, any $\varphi \in (0, \pi/2)$ exhibits a certain level of altruism. We adopt only the values of these definitions for the purposes of illustration and numerical study.

\subsection{Socially Compliant Control of AVs with Application to Eco-driving}\label{section3.2}

We present a general formulation of an optimization problem for socially compliant control design of AVs using SVO introduced above, followed by a specific example considering eco-driving of AVs. Following the definition shown in equation~\eqref{eq3.1} we consider the interactions between a controlled AV and the vehicle immediately behind in designing socially compliant autonomous driving policy. In other words, by considering the impact on the following vehicle the controlled AV is expected to drive in a socially compliant manner, thereby enabling the follower to better understand and appropriately respond to its actions. For any controlled AV, the utility function $J_1$ shown in equation~\eqref{eq3.1} is to be maximized
\begin{eqnarray}
    \max J_1 = \max \left( \cos{(\varphi_{\text{AV}})} U_{\text{AV}} + \sin{(\varphi_{\text{AV}})} U_{\text{follower}} \right),    \label{eq3.2}
\end{eqnarray}
under a set of constraints on the operation of the AV, such as control constraints on the additive input variable $u_i$ appearing in equation~\eqref{eq2.5b}. Clearly, equation~\eqref{eq3.2} covers a broad class of optimization problems for AV control design, depending on the construction of the individual payoff functions, $U_{\text{AV}}$ and $U_{\text{follower}}$. 

In what follows, we apply the concept presented above to design socially compliant eco-driving controls for AVs. It is worth noting that prior studies on eco-driving controls of AVs have largely focused on minimizing the energy consumption of controlled AVs, e.g.,~\citep{sun2022energy,wang2019cooperative,yang2020eco}, without explicitly considering the impact of their driving behavior on the following vehicle in the control design process. By contrast, we incorporate SVO into eco-driving control design for AVs, where the objective is to synthesize an energy-efficient, yet socially compliant, driving strategy for AVs. As a result, the following vehicle, particularly when being human-driven, would be able to better comprehend the actions of the AV immediately ahead and responds appropriately. To this end, we shall construct explicit forms of the payoff functions $U_{\text{AV}}$ and $U_{\text{follower}}$.  

$U_{\text{AV}}$ is related to the energy consumption of the AV considering the goal of eco-driving. Following~\citep{zhang2016optimal,malikopoulos2018decentralized}, we aim to minimize the effective acceleration input, i.e., $\dot{v}_i$, $i \in \mathcal{A}$, of the AV to reduce its energy cost. In other words, we minimize transient engine operation, leading to direct improvements in fuel consumption. This is because internal combustion engines are designed for optimal performance under steady-state operating conditions, which involve consistent torque and speed~\citep{malikopoulos2018decentralized}. Consequently, one can formulate $U_{\text{AV}}$ as follows
\begin{eqnarray}
    U_{\text{AV}} = - \int_{I} (1/2)\dot{v}_i(t)^2 dt, ~i \in \mathcal{A},    \label{eq3.3}
\end{eqnarray}
where $I = \left[t_0, t_f\right]$ denotes the operation horizon. It is noted that $U_{\text{AV}}$ given by equation~\eqref{eq3.3} is non-positive. In other words, a larger effective acceleration input, or equivalently greater energy consumption, would result in a smaller payoff for the controlled AV. This is consistent with the eco-driving objective of minimizing vehicle fuel consumption. 

For any $(i+1) \in \mathcal{H}$ following the AV, $U_{\text{follower}}$ may have non-unique characterizations, depending on the objective of the human driver behind the controlled AV. For instance, drivers tend to achieve a desired speed which is normally considered as the speed limit or free-flow speed~\citep{treiber2000congested}. Hence, the $U_{\text{follower}}$ associated with this common case can be defined as
\begin{eqnarray}
    U_{\text{follower}} = - \int_{I} (1/2) \left[v_{i+1}(t) - v_0\right]^2 dt, ~i+1 \in \mathcal{H},    \label{eq3.4}
\end{eqnarray}
where $v_0$ signifies the speed limit for road traffic. 

Plugging equations~\eqref{eq3.3} and~\eqref{eq3.4} into the expression of~\eqref{eq3.2} yields
\begin{eqnarray}
    \max J_1 = -\max \left( \cos{(\varphi_{\text{AV}})} \int_{I} (1/2)\dot{v}_i(t)^2 dt + \sin{(\varphi_{\text{AV}})} \int_{I} (1/2) \left[v_{i+1}(t) - v_0\right]^2 dt \right),  \label{eq3.5}
\end{eqnarray}
which is equivalent to 
\begin{eqnarray}
    \max J_1 = \min J_2 = \min \left( \cos{(\varphi_{\text{AV}})} \int_{I} (1/2)\dot{v}_i(t)^2 dt + \sin{(\varphi_{\text{AV}})} \int_{I} (1/2) \left[v_{i+1}(t) - v_0\right]^2 dt \right),  \label{eq3.6}
\end{eqnarray}
where $J_2$ is as defined as the right-hand side of equation~\eqref{eq3.6}. 

It is worthwhile to note that so far we have considered a fundamental scenario with no vehicle in the vicinity ahead of the controlled AV. While such a setting is important, it is also likely that the design of AV driving strategies could be constrained by the presence of an immediate preceding vehicle. As seen in~\citep{sun2022energy}, it is useful to ensure a reasonable traffic throughput while deriving eco-driving policies for an AV. Hence, we introduce a soft state constraint given by 
\begin{eqnarray}
    s_i = s_{\text{d}},    \label{eq3.7}
\end{eqnarray} 
where $s_i$ is the inter-vehicle spacing between the controlled AV and the vehicle immediately ahead (which can be easily obtained using onboard sensors~\citep{wang2022optimal,wang2023general}), and $s_{\text{d}} (\geq s_0)$ signifies the desired spacing which may not necessarily be achieved. It is noted that one can choose a larger value of $s_{\text{d}}$ to characterize more conservative driving behavior of an AV. Now we introduce an additional positive Lagrange multiplier $\lambda$ and redefine the objective functional $J_2$ as
\begin{eqnarray}
    J_3 = \int_{I} (1/2)\left\{ \cos{(\varphi_{\text{AV}})}\dot{v}_i(t)^2 + \sin{(\varphi_{\text{AV}})}\left[v_{i+1}(t) - v_0\right]^2 + \lambda\left[s_i(t) - s_{\text{d}}\right]^2 \right\}dt,    \label{eq3.8}
\end{eqnarray} 
which is to be minimized. It is noted that safety of the AV can be ensured since the optimization tends to minimize AV acceleration and one can choose a large value of $s_{\text{d}}$ or $\lambda$ to achieve a desired safe spacing.

\begin{remark}\label{Remark3.1}
While equation~\eqref{eq3.4} presents one reasonable formulation of the payoff function for HVs, there are other choices possible depending on the objective of the human driver. For instance, one can define $U_{\text{follower}} = \int_{I} (1/2) \left[v_{i+1}(t)\right]^2 dt, ~i+1 \in \mathcal{H}$, if the HV prefers to travel as fast as possible; or $U_{\text{follower}} = -\int_{I} (1/2) \left[v_{i+1}(t)-v_i(t)\right]^2 dt$, if the HV tends to drive in a smoother manner~\citep{wang2023general,wang2022smoothing}.  
\end{remark}

\subsection{Numerical Solution}\label{section3.3}

We consider a dynamic system of two-vehicle car following with the controlled AV being followed by a HV for design of socially compliant AV controls in the context of eco-driving. As mentioned before, we focus on human-robot interactions and hence do not consider the setting where an AV is followed by another AV. This is reasonable since socially compliant control design is expected to be carried out with regard to human-AV interaction, as seen in~\citep{schwarting2019social}. Thus, for vehicle $i \in \mathcal{A}$ followed by $(i+1) \in \mathcal{H}$, we define the system state vector $y = \left[ x_{i}, v_i, x_{i+1}, v_{i+1} \right]^{T}$ and the control input $u = u_{i}$, where the superscript $T$ denotes the transpose operator.

We now consider the optimal control problem, i.e., a Lagrange problem, minimizing the objective functional $J_3$ reproduced below in a compact form
\begin{eqnarray}
    J_3 \coloneqq \int_{I} \ell(t,y,u)dt,    \label{eq3.9}
\end{eqnarray} 
where $\ell$ is equivalent to the integrand shown on the right-hand side of equation~\eqref{eq3.8}. The abstract notation $\ell$ is introduced for ease of mathematical analysis. 

The time derivative of $y$ is given by
\begin{eqnarray}
\dot{y} =
\begin{bmatrix} 
v_i \\ \dot{v}_{i} \\ v_{i+1} \\ \dot{v}_{i+1}  
\end{bmatrix}
\coloneqq g(t,y,u),
\label{eq3.10}
\end{eqnarray} 
where $\dot{v}_{i}$ and $\dot{v}_{i+1}$ are computed using equations~\eqref{eq2.5b} and~\eqref{eq2.5a}, respectively, and $g$ is a vector field representing system dynamics. It is worth noting that the state variables, i.e., position and speed, of the vehicle immediately ahead of the controlled AV are not explicitly written out, since this information is readily accessible to the AV via its onboard sensors~\citep{wang2022optimal,wang2023general}. The additive acceleration input $u$ takes values from the following control constraint set 
\begin{eqnarray}
U \coloneqq \left\{ u: u_{\text{min}} \leq u \leq u_{\text{max}} \right\}.   \label{eq3.11}
\end{eqnarray} 

We now apply Pontryagin's minimum principle (PMP)~\citep{pontryagin1962mathematical} to solve the Lagrange problem formulated above, with the objective functional defined in equation~\eqref{eq3.9}, system dynamics shown in equation~\eqref{eq3.10}, and the control constraint set given by equation~\eqref{eq3.11}. Optimal control theory has been well established and extensively employed in various mathematical and engineering disciplines, e.g., see~\citep{ahmed2021optimal} and the references therein. The classic optimal control method, namely PMP~\citep{pontryagin1962mathematical}, has been adopted for solving a series of engineering problems such as stabilizing building maintenance units~\citep{wang2021optimalJIMO}, allocating parking space for AVs~\citep{wang2021optimal}, integrating AVs into the auto market~\citep{wang2022policy}, among others. Essentially, the PMP prescribes a mathematically proven optimal control policy that drives a dynamic system to the desired state. To this end, we shall first define a Hamiltonian function $H$ as follows
\begin{eqnarray}
H(t,y,u,\psi) \coloneqq \langle g(t,y,u),\psi \rangle + \ell(t,y,u),    \label{eq3.12}
\end{eqnarray}
where $\psi = \left[\psi_{1},\psi_{2},\psi_{3},\psi_{4}\right]^{T}$ signifies the adjoint state vector and the operation $\langle A, B \rangle$ represents the inner product of the two vectors, $A$ and $B$. Following the Hamiltonian function defined in~\eqref{eq3.12}, the adjoint state equation is given by
\begin{eqnarray}
\dot{\psi} = -\frac{\partial H}{\partial y} = -g_y^T(t, y, u)\psi - \ell_y(t,y,u), ~t \in I,     \label{eq3.13}
\end{eqnarray}
where $g_y$ and $\ell_y$ denote the Jacobian matrices of $g$ and $\ell$, respectively, and $g_y^T$ is the transpose of $g_y$. The boundary condition on the adjoint state $\psi$ is given by $\psi(t_f) = 0$ at the time $t_f$ due to absence of a terminal cost in the objective functional~\eqref{eq3.9}. Consequently, the following necessary conditions of optimality hold
\begin{eqnarray}
H(t,y^{o}(t),u^{o}(t),\psi^{o}(t)) \leq H(t,y^{o}(t),u(t),\psi^{o}(t)),    \label{eq3.14}
\end{eqnarray}
subject to the equations~\eqref{eq3.10} and~\eqref{eq3.13} of the state and adjoint state, respectively. The superscript $o$ indicates the optimal value. 

As a result, the partial derivative of the Hamiltonian concerning the additive control input is given by
\begin{eqnarray}
H_u = \frac{\partial H}{\partial u} =  \psi_2 + \cos{(\varphi_{\text{AV}})} \left[ k_1(x_{\text{p}} - y_1 - l_{\text{p}} - \eta - \tau_2 y_2) + k_2(v_{\text{p}} - y_2) + u \right],
\label{eq3.15}
\end{eqnarray}
where $x_{\text{p}}$, $v_{\text{p}}$, and $l_{\text{p}}$ denote the position, speed, and length of the vehicle immediately ahead of the controlled AV. The partial derivative $H_u$ allows one to numerically compute the solution of the formulated optimal control problem in an iterative manner. Equations~\eqref{eq3.10} and~\eqref{eq3.13} are solved forward and backward in time, respectively. The iterative computational procedure is derived following a gradient descent method that generates a sequence of control policies $\{u^{(\kappa)}\}$, along which the objective functional $J_3$ converges to its minimum. Following each iteration, equation~\eqref{eq3.15} provides the optimal descent direction for constructing the control policy in the subsequent iteration, e.g., see Step 6 below. The primary numerical procedures are succinctly outlined as follows:

\vskip4pt 
Step 1: Initialization. Set the initial state of the car-following system, the first and maximum iteration numbers as $\kappa=1$ and $N_{\text{max}}=300$, respectively, and other system parameter values including the SVO angular preference $\varphi_{\text{AV}}$ of AVs.

Step 2: Integrate the state equation~\eqref{eq3.10} forward in time, and record the state trajectory $y^{(\kappa)}$ and the control input vector $u^{(\kappa)}$.

Step 3: Integrate the adjoint equation~\eqref{eq3.13} backward in time, and record the trajectory of the adjoint state $\psi^{(\kappa)}$.

Step 4: Use the triple \{$y^{(\kappa)}, u^{(\kappa)}, \psi^{(\kappa)}$\} to compute the gradient $H_u$ using equation~\eqref{eq3.15}, and calculate the objective functional $J_3^{(\kappa)}$ according to equation~\eqref{eq3.8}.

Step 5: If the square of $H_u$ is smaller than a specified small positive value, say $\delta$, then set $u^o = u^{(\kappa)}$ and $J_3^o = J_3^{(\kappa)}$; otherwise, go to Step 6.

Step 6: Use $u^{(\kappa)}$ to construct the subsequent control policy, $u^{(\kappa+1)} = u^{(\kappa)} - \epsilon H_u(t, y^{(\kappa)}, u^{(\kappa)}, \psi^{(\kappa)})$, with a sufficiently small step size $\epsilon \in (0,1)$ such that $u^{(\kappa+1)} \in U$. Specifically, if $u^{(\kappa+1)} > u_{\text{max}}$ let $u^{(\kappa+1)} = u_{\text{max}}$, and let $u^{(\kappa+1)} = u_{\text{min}}$ if $u^{(\kappa+1)} < u_{\text{min}}$. By applying the Lagrange formula, it is readily verified that the objective functional $J_3$ at $u^{(\kappa+1)}$ can be expressed in terms of $J_3$ at $u^{(\kappa)}$ as follows
\begin{eqnarray}
&~&\hskip-30pt J_3(u^{(\kappa+1)}) = J_3(u^{(\kappa)}) - \epsilon \left|H_u(t, y^{(\kappa)}, u^{(\kappa)}, \psi^{(\kappa)})\right|^2 + o(\epsilon),  \nonumber \\   \label{eq3.16}
\end{eqnarray}
where $o(\epsilon)$ is the higher order terms. Hence, $J_3(u^{(\kappa+1)}) < J_3(u^{(\kappa)})$ for $\epsilon > 0$ sufficiently small.

Step 7: Compute the value of $J(u^{(\kappa+1)})$ and check if 
\begin{eqnarray}
\left|J_3(u^{(\kappa+1)}) - J_3(u^{(\kappa)})\right| > \Upsilon   \label{eq3.17}
\end{eqnarray}
holds for a prespecified small positive value $\Upsilon \in \mathbb{R}^+$. If so, go to Step 2 with $u^{(\kappa)}$ replaced by $u^{(\kappa+1)}$ and repeat the process. For the predefined $\Upsilon > 0$, the procedure is continued while $\left|J_3(u^{(\kappa+1)}) - J_3(u^{(\kappa)})\right| > \Upsilon$ holds before reaching the maximum number of iterations, $N_{\text{max}}$. Otherwise, the computational procedure is terminated. 

\begin{remark}\label{Remark3.2}
From the preceding analysis it is clear that, for any feasible initial selection of $u^{(1)}$ the algorithm presented produces a series of AV control inputs, $\{u^{(\kappa)}\}$, along which the objective functional $J_3(u^{(\kappa)})$ decreases monotonically, provided a suitably small step size $\epsilon$ is used, i.e., $J_3(u^{(1)}) > J_3(u^{(2)}) > J_3(u^{(3)}) > \cdots$. Hence, there exists a real number $\alpha$ such that $\lim_{\kappa \rightarrow \infty} J_3(u^{(\kappa)}) \longrightarrow \alpha$. Therefore, the solution converges potentially to a local minimum since a nonlinear optimal control problem is being considered. More general convergence results on measure-valued solutions of dynamic systems can be found in~\citep{ahmed2021optimal}.
\end{remark}

\section{Numerical Results}\label{section4}

\begin{table}[t!]
\normalfont
\footnotesize
\setlength{\tabcolsep}{2pt}
\caption{\textnormal{Model Parameter Values of the IDM and OVRV Model}}\label{Table_parameters}
\begin{center}
 \begin{tabular}{c c c c}
 \hline \hline
 \textbf{Parameter} ~&~ \textbf{Description} ~&~ \textbf{IDM} ~&~ \textbf{OVRV}  \\  [0.5ex]
 \hline
 $v_{0}$ ~&~ desired speed (\text{m/s}) ~&~ 30.0 ~&~ - \\ [0.3ex]
 \hline
 $\tau_1$ ~&~ time gap (\text{s}) ~&~ 1.5 ~&~ - \\  [0.3ex]
 \hline
 $s_{0}$ ~&~ minimum spacing (\text{m}) ~&~ 2.0 ~&~ -\\[0.3ex]
 \hline
 $a$ ~&~ maximum acceleration (\text{m/s\textsuperscript{2}}) ~&~ 1.0 ~&~ -  \\  [0.3ex]
 \hline
 $b$ ~&~ comfortable deceleration (\text{m/s\textsuperscript{2}}) ~&~ 1.5 ~&~ - \\  [0.3ex]
 \hline
 $l_{i}$ ~&~ vehicle length (\text{m}) ~&~ 5.0 ~&~ 5.0  \\ [0.3ex]
 \hline
 $k_{1}$ ~&~ gain parameter on the time gap (\text{s\textsuperscript{-2}}) ~&~ - ~&~ 0.1   \\  [0.3ex]
 \hline
 $k_{2}$ ~&~ gain parameter on the relative speed (\text{s\textsuperscript{-1}}) ~&~ - ~&~ 0.6  \\  [0.3ex]
 \hline
 $\eta$ ~&~ jam distance (\text{m}) ~&~ - ~&~ 21.51  \\  [0.3ex]
 \hline
 $\tau_2$ ~&~ desired time gap (\text{s}) ~&~ - ~&~ 1.71  \\ [0.3ex]
 \hline \hline
\end{tabular}
\end{center}
\end{table}

\begin{figure}[t!]
	\centering
	\includegraphics[width=\textwidth]{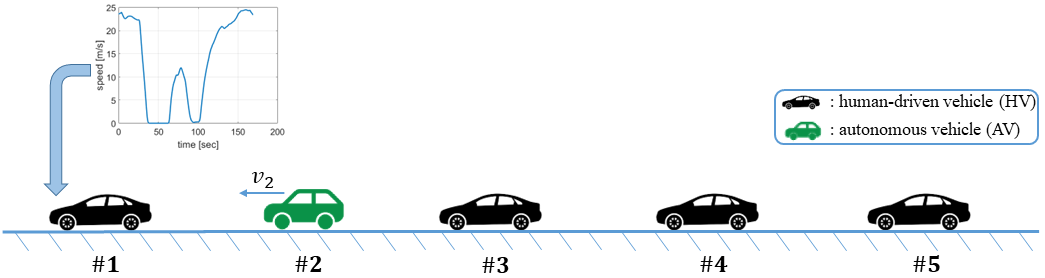}
	\caption{Graphic illustration of a string of five vehicles, with the lead human-driven vehicle (HV) executing a real-world speed profile acquired on Highway~55 in Minnesota~\citep{sun2022energy}. The lead vehicle slows down at two consecutive intersections due to red signals.}
	\label{Simulation_setting}
\end{figure}

In this section, we carry out a set of numerical simulations using~\texttt{MATLAB} to illustrate the mechanism and efficacy of the developed methodology. Following Section~\ref{section2}, the dynamics of HVs and AVs are described by the IDM and OVRV model, respectively, with their corresponding parameter values~\citep{treiber2013traffic,gunter2019modeling} summarized in Table~\ref{Table_parameters}. The proposed control design method works independently for each individual AV without necessitating cooperation or communication between vehicles, i.e., in a car-following context involving the AV and its immediate follower. As shown in Fig.~\ref{Simulation_setting}, we consider a generic setting where the controlled AV (\#2) follows a HV (\#1) executing a real-world speed profile acquired on Highway~55 in Minnesota~\citep{sun2022energy}. As mentioned before, we consider a general case of human-AV interactions with the controlled AV being followed by a HV (\#3). To examine how the driving behavior of the AV affects the following vehicles, we consider a string of 3 HVs (\#3, \#4, \#5) driven behind the AV, while more vehicles could be included. Since the proposed AV driving strategy does not require vehicle communication, it can be effectively applied to individual AVs without depending on their market penetration rate, i.e., regardless of the number of AVs within the traffic flow. As mentioned before, we focus exclusively on the longitudinal vehicle dynamics, similar to previous research~\citep{wang2022optimal,wang2023general,talebpour2016influence}, while lateral dynamics could also be explored. 

In numerical simulation, the constraint set for the additive control input is given by $U = \left\{ u: -0.6 \leq u \leq 0.6 \right\}$, and the parameter values, i.e., Lagrange multiplier $\lambda = 0.01$, desired spacing $s_{\text{d}} = 10$~m, and step size $\epsilon = 0.01$, are used for illustrative purposes. The three SVO values of the AV are given by $\varphi_{\text{AV}} = \pi/2$, $\pi/4$, and 0.1 to represent altruistic, prosocial, and egoistic preferences, respectively. It is worth noting that a small angular value of $\varphi_{\text{AV}} = 0.1$ is used, instead of 0, to denote the egoistic preference of the AV. This is to balance the three terms on the right-hand side of equation~\eqref{eq3.8} for a reasonable solution to the optimal control problem, since the first term could be significantly smaller than the others that are on a similar scale. In what follows, we study the effectiveness of the proposed method and the impact of its resulting AV driving strategy on the following vehicles, considering different social preferences of the controlled AV. 

\begin{figure}[t!]
	\centering
	\subfloat[$\varphi_{\text{AV}} = \pi/2$]{\includegraphics[width=0.33\textwidth]{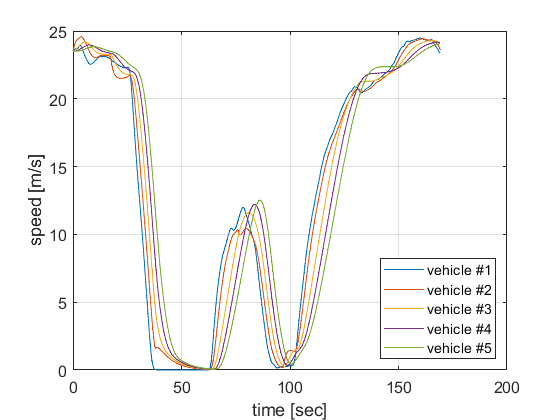}\label{speed_1}}
	\hfil%
	\subfloat[$\varphi_{\text{AV}} = \pi/4$]{\includegraphics[width=0.33\textwidth]{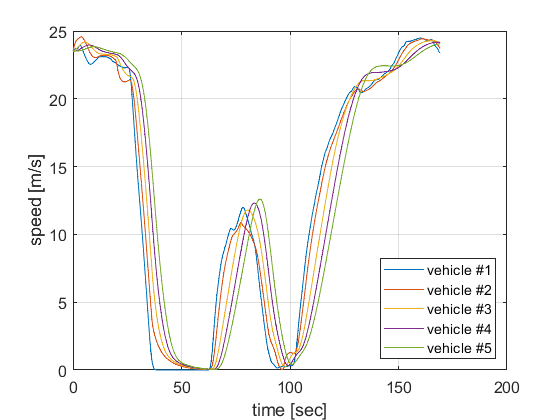}\label{speed_2}}
	\hfil%
	\subfloat[$\varphi_{\text{AV}} = 0.1$]{\includegraphics[width=0.33\textwidth]{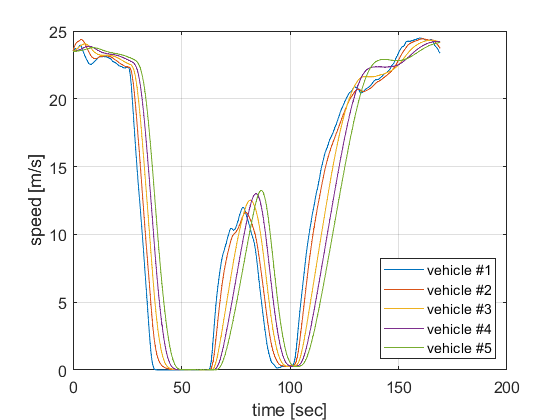}\label{speed_3}} 
	\caption{\textnormal{Speed profile of all vehicles considering different social preferences of AVs}}\label{speed_set_1}
\end{figure}

\begin{figure}[t!]
	\centering
	\subfloat[$\varphi_{\text{AV}} = \pi/2$]{\includegraphics[width=0.33\textwidth]{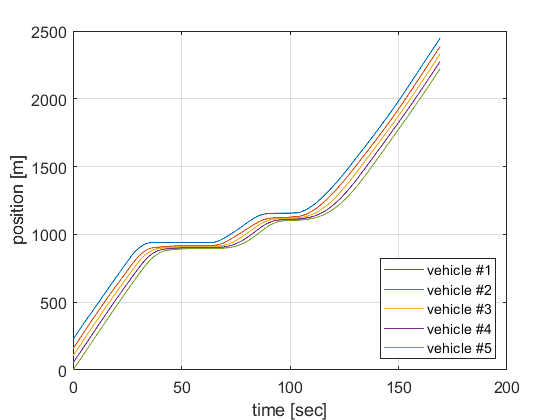}\label{position_1}}
	\hfil%
	\subfloat[$\varphi_{\text{AV}} = \pi/4$]{\includegraphics[width=0.33\textwidth]{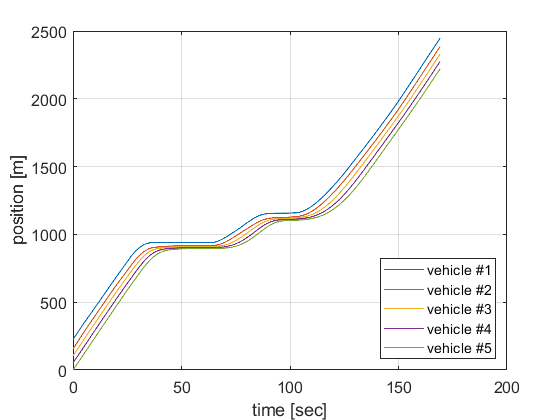}\label{position_2}}
	\hfil%
	\subfloat[$\varphi_{\text{AV}} = 0.1$]{\includegraphics[width=0.33\textwidth]{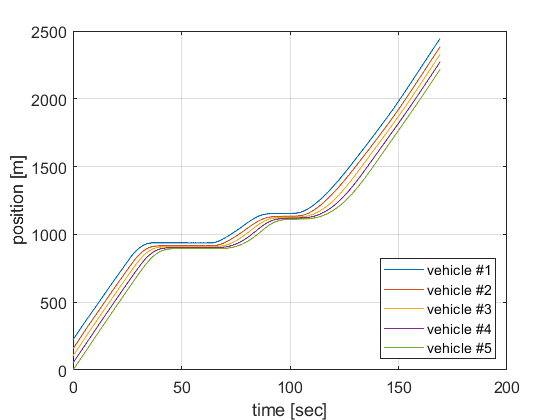}\label{position_3}} 
	\caption{\textnormal{Trajectory profile of all vehicles considering different social preferences of AVs}}\label{position_set_1}
\end{figure}

Fig.~\ref{speed_set_1} and Fig.~\ref{position_set_1} show the speed and trajectory profiles of all the vehicles for $\varphi_{\text{AV}} = \pi/2$, $\pi/4$, and 0.1. It is observed from Fig.~\ref{speed_set_1} that, starting from the controlled AV (\#2) vehicles exhibit a faster slowdown at red signals as the value of $\varphi_{\text{AV}}$ decreases, i.e., $\pi/2 > \pi/4 > 0.1$, aiming to keep their speed further away from the desired value of $v_0$. This is due to the fact that an AV with a more egoistic driving preference (smaller value of $\varphi_{\text{AV}}$) tends to minimize its own energy consumption at the expense of sacrificing the follower's payoff. A set of comprehensive results are summarized in Table~\ref{Table_evaluation}, where a larger value of $\left|U_{\text{AV}}\right|$ indicates greater energy consumption of the AV, or equivalently a smaller utility due to the negative sign in equation~\eqref{eq3.3}. It is observed from Table~\ref{Table_evaluation} that, with a more altruistic preference (larger value of $\varphi_{\text{AV}}$) the controlled AV can help increase the average speed of the following vehicles (equivalent to decreasing their travel times) at the expense of greater energy consumption, given that the payoffs of the AV and HVs are to reduce fuel consumption and achieve the desired speed $v_0$, respectively.

Specifically, from being egoistic ($\varphi_{\text{AV}} = 0.1$) to prosocial ($\varphi_{\text{AV}} = \pi/4$) and altruistic ($\varphi_{\text{AV}} = \pi/2$), the value of $\left|U_{\text{AV}}\right|$ for the AV increases by 5.00\% and 7.21\%, respectively, as shown in Table~\ref{Table_evaluation}. Consequently, the average speeds of the following vehicles experience corresponding increases of 0.46\% and 6.47\% (for~\#3), 0.26\% and 5.98\% (for~\#4), 0.27\% and 5.47\% (for~\#5), respectively. It is interesting to note that a following vehicle closer to the controlled AV generally tends to obtain greater benefits thanks to the courtesy of the AV using socially compliant controls, consistent with the findings of~\citep{sun2022energy,wang2023general}.

\begin{table}[t!]
\caption{\textnormal{Evaluation of the absolute payoff $\left|U_{\text{AV}}\right|$ of the AV and the average speed of HVs following the AV, with the simulation settings shown in Fig.~\ref{Simulation_setting}. Following equation~\eqref{eq3.3}, a larger value of $\left|U_{\text{AV}}\right|$ (i.e., a smaller utility for the controlled AV) corresponds to greater acceleration input, indicating higher energy consumption for internal combustion engine vehicles as shown in~\citep{malikopoulos2018decentralized}.}}\label{Table_evaluation}
\begin{center}
 \begin{tabular}{c|c c c}
 \textbf{Metric} ~&~ \textnormal{$\varphi_{\text{AV}} = \pi/2$} ~&~ \textnormal{$\varphi_{\text{AV}} = \pi/4$} ~&~ \textnormal{$\varphi_{\text{AV}} = 0.1$} \\  [0.5ex]
 \hline
 \textnormal{$\left|U_{\text{AV}}\right|$} ~&~ \textnormal{477.6331} ~&~ \textnormal{467.8203} ~&~ \textnormal{445.5223} \\  [0.3ex]
 \hline
 \textnormal{Average speed of vehicle \#3 (m/s)} ~&~ \textnormal{10.444} ~&~ \textnormal{9.854} ~&~ \textnormal{9.809}  \\  [0.3ex]
 \hline
 \textnormal{Average speed of vehicle \#4 (m/s)} ~&~ \textnormal{10.629} ~&~ \textnormal{10.055} ~&~ \textnormal{10.029} \\  [0.3ex]
 \hline
 \textnormal{Average speed of vehicle \#5 (m/s)} ~&~ \textnormal{10.888} ~&~ \textnormal{10.351} ~&~ \textnormal{10.323} \\  [0.3ex]
 \hline 
\end{tabular}
\end{center}
\end{table}

\begin{figure}[t!]
	\centering
	\includegraphics[width=0.4\textwidth]{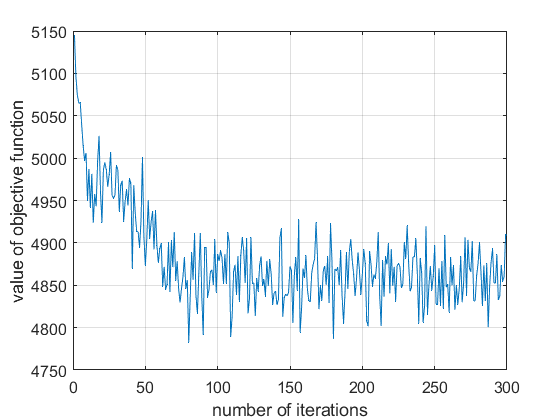}
	\caption{\textnormal{Value of the objective functional $J_3$ corresponding to $\varphi_{\text{AV}} = 0.1$}}
	\label{objective_function}
\end{figure}

To show convergence of computation, the objective functional value, $J_3$, corresponding to $\varphi_{\text{AV}} = 0.1$, is presented in Fig.~\ref{objective_function}. Clearly, as the number of iterations increases, the value of $J_3$ tends to decrease and converge, indicating convergence of the computational algorithm. It is worth noting that $J_3$ does not exhibit a monotonic decrease. This behavior is a result of choosing a step size $\epsilon$ that is not adequately small to facilitate faster convergence, and a discussion on this aspect can be found in~\cite[Chapter~7]{ahmed2021optimal}. The numerical value to which $J_3$ converges depends on its formulation shown in equation~\eqref{eq3.8}. 

\begin{figure}[t!]
	\centering
	\includegraphics[width=0.5\textwidth]{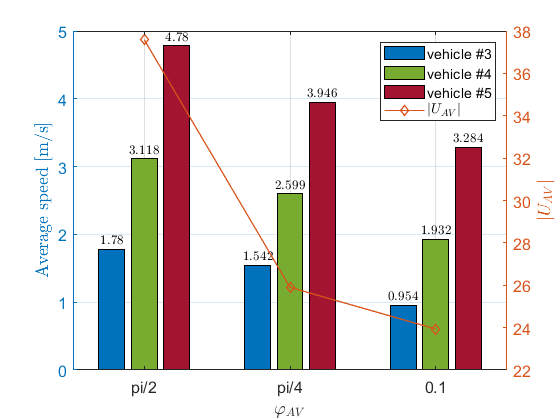}
	\caption{\textnormal{Evaluation of the absolute payoff $\left|U_{\text{AV}}\right|$ of the AV and the average speed of HVs following the AV, over the period of $[30, 60]$~sec. According to equation~\eqref{eq3.3}, a larger value of $\left|U_{\text{AV}}\right|$ signifies a lower utility for the controlled AV due to the negative sign, indicating higher energy consumption.}}
	\label{evaluation_segment}
\end{figure}

It is observed from Fig.~\ref{speed_1}--\ref{speed_3} that the results do not show significant differences, resulting in relatively small numerical improvements in average speed for the HVs following the AV over the entire simulation horizon (Table~\ref{Table_evaluation}). However, it is important to note that the effect of eco-driving becomes more pronounced as the controlled AV approaches a signalized intersection. To this end, we present a set of results corresponding to the time period of $[30, 60]$ seconds in Fig.~\ref{evaluation_segment}, during which vehicles have to slow down due to a red light. It is observed from Fig.~\ref{evaluation_segment} that, from being egoistic ($\varphi_{\text{AV}} = 0.1$) to prosocial ($\varphi_{\text{AV}} = \pi/4$) and altruistic ($\varphi_{\text{AV}} = \pi/2$), the value of $\left|U_{\text{AV}}\right|$ for the AV increases by 8.21\% and 57.04\%, respectively. Consequently, the average speeds of the following vehicles experience corresponding increases of 61.64\% and 86.58\% (for~\#3), 34.52\% and 61.39\% (for~\#4), 20.16\% and 45.55\% (for~\#5), respectively. These results of percentage change in $\left|U_{\text{AV}}\right|$ and average speed of the following vehicles are summarized in Table~\ref{Table_evaluation_segment}. Clearly, a following vehicle closer to the controlled AV tends to obtain greater benefits thanks to the courtesy of the AV using socially compliant controls, e.g., 61.64\% $>$ 34.52\% $>$ 20.16\% with $\varphi_{\text{AV}} = \pi/4$. These results are also consistent with the recent findings of~\citep{sun2022energy,wang2023general}.

\begin{table}[t!]
\caption{\textnormal{Evaluation of the percentage change in the utility of the AV ($\left|U_{\text{AV}}\right|$) and the average speed of HVs following the AV over the time period of $[30, 60]$ seconds, during which vehicles have to slow down due to a red light. It is worth noting that all the percentage changes are calculated with the values for $\varphi_{\text{AV}} = 0.1$ (egoistic) serving as the base for comparison.}}\label{Table_evaluation_segment}
\begin{center}
 \begin{tabular}{c|c c c}
 \textbf{Metric (percentage change)} ~&~ \textnormal{$\varphi_{\text{AV}} = 0.1$ (egoistic)} ~&~ \textnormal{$\varphi_{\text{AV}} = \pi/4$ (prosocial)} ~&~ \textnormal{$\varphi_{\text{AV}} = \pi/2$ (altruistic)} \\  [0.5ex]
 \hline
 \textnormal{$\left|U_{\text{AV}}\right|$} ~&~ \textnormal{--} ~&~ \textnormal{+8.21\%} ~&~ \textnormal{+57.04\%} \\  [0.3ex]
 \hline
 \textnormal{Average speed of vehicle \#3 (m/s)} ~&~ \textnormal{--} ~&~ \textnormal{+61.64\%} ~&~ \textnormal{+86.58\%}  \\  [0.3ex]
 \hline
 \textnormal{Average speed of vehicle \#4 (m/s)} ~&~ \textnormal{--} ~&~ \textnormal{+34.52\%} ~&~ \textnormal{+61.39\%} \\  [0.3ex]
 \hline
 \textnormal{Average speed of vehicle \#5 (m/s)} ~&~ \textnormal{--} ~&~ \textnormal{+20.16\%} ~&~ \textnormal{+45.55\%} \\  [0.3ex]
 \hline 
\end{tabular}
\end{center}
\end{table}

\begin{remark}\label{Remark4.1}
To reduce complexity of implementing the proposed method, the state constraints for car following are implicitly considered in the presented framework in a conservative fashion, as also seen in~\citep{wang2022optimal}. Specifically, one shall select a control constraint set $U$ with a limited range to ensure the maintenance of a desired minimum following distance. To that end, a Lagrange multiplier $\lambda$ is introduced into the objective functional to construct an augmented one, as seen in Section~\ref{section3.2}. As an alternative, the constraint transcription technique~\citep{teo1991unified} could be used to further improve the numerical results.
\end{remark}

\section{Conclusions and Future Work}\label{section5}

In this study, we have established a general framework for socially compliant AV control design using social value orientation (SVO), a metric from social psychology. SVO enables AVs to influence the behavior of following HVs. Within this framework, we maximize the utilities of the controlled AV and its follower, considering the AV's SVO. This optimization can address various design objectives, aligning with the AV's goal and the benefits for the immediate follower due to socially compliant AV control. We formulate an optimal control problem to maximize the utility function and solve it numerically using Pontryagin's minimum principle, ensuring optimality. We apply this methodology to synthesize socially compliant control for eco-driving AVs. Real-world vehicle trajectory data in numerical simulations demonstrate the effectiveness of our approach.

While the results presented are promising for socially compliant AV control design in mixed traffic, there is room for further study and extensions in the future. A few interesting research directions for future studies are listed as follows.
\begin{itemize}
    \item Explore different design objectives as mentioned in Remark~\ref{Remark3.1}, and assess the effectiveness of the proposed approach. This is important because personal AVs may have varying operational goals depending on their owners~\citep{shang2023capacity}.
    
    \item Consider expanding the study beyond longitudinal dynamics for AV control design. While this aligns with existing work, recent research has shown that additional energy benefits can be achieved by considering lane-changing impacts on eco-driving controls~\citep{he2023real}. Thus, it is worthwhile to investigate two-dimensional vehicular movement in control design for future studies.
        
    \item Although the proposed AV control approach does not require vehicle communication, vehicular connectivity is likely to be part of future transportation systems. Therefore, it is useful to explore the cooperation among AVs in the context of connectivity topologies for socially compliant cooperative control design. 
\end{itemize}

\bibliographystyle{unsrtnat}
\bibliography{arxiv}  

\end{document}